# Effective conductivity in a lattice model for binary disordered media with complex distributions of grain sizes

R. Piasecki[*]

*Institute of Chemistry, University of Opole, Oleska 48, 45-052 Opole, Poland*

Using numerical simulations and analytical approximations we study a modified version of the two-dimensional lattice model [R. Piasecki, phys. stat. sol. (b) **209**, 403 (1998)] for random $p\text{H}:(1-p)\text{L}$ systems consisting of grains of high (low) conductivity for the H-(L-)phase, respectively. The modification reduces a spectrum of model bond conductivities to the two pure ones and the mixed one. The latter value explicitly depends on the average concentration $\chi(p)$ of the H-component per model cell. The effective conductivity as a function of content $p$ of the H-phase in such systems can be modelled making use of three model parameters that are sensitive to both grain size distributions, GSD(H) and GSD(L). However, to incorporate into the model information directly connected with a given GSD, a computer simulation of the geometrical arrangement of grains is necessary. By controlling the polydispersity in grain sizes and their relative area frequencies, the effective conductivity could be raised or decreased and correlated with $\chi(p)$. When the phases are interchanged, a hysteresis-loop like behaviour of the effective conductivity, characteristic of dual media, is found. We also show that the topological non-equivalence of system's microstructure accompanies some GSDs, and it can be detected by the entropic measure of spatial inhomogeneity of model cells.

## 1. Introduction

The effective properties of binary disordered media are often predicted by simple models on the basis of the component properties (see [1] for a recent review). Both theoretical and experimental investigations devoted to particle size influence on macroscopic properties attract increasing attention. Incorporation of grain size distributions (GSDs) into the effective medium approach (EMA) by a two-dimensional lattice model [2] for the effective conductivity of binary granular systems was motivated by recent Monte Carlo simulations [3, 4]. The latter, though performed for mono-sized grains, revealed oscillations of the critical percolation coverage as a function of the size of grains. For finite-sized objects with circular symmetry in two dimensions and spherical one in three dimensions, the percolation threshold is a well-defined quantity as shown by numerical investigations on a lattice [5]. On the other hand, the effect of the geometrical arrangement of equally sized hard 'spheres' on the average electrical properties of two- and three-component composites was simulated using a modified transfer matrix algorithm for a triangular network [6]. The results compared with some common mixture formulas indicate that the geometrical arrangement effects become more important as the sample size decreases and the filling fraction approaches the effective percolation threshold. It was assumed there that the spheres fill the space in a close-packed pattern.

In this context, it is worth to mention the experimental studies [7] how several shapes composed of welded spheres (doubles, hexagons, small and large triangles, diamonds, triples, trapezoids) pack in two dimensions. Ordered domains occurred near a packing density of 0.8 except trapezoids, which all packings remain disordered and near the transition density, even after annealing by shaking. A numerical work on mono-sized square and rod-like particles was done for the description of the conductivity of more specified materials, solid electrolytes [8]. A complementary study within EMA in both two and three dimensions for dispersed ionic conductors revealed analytical expressions for the particle size dependence of the percolation thresholds of the model [9]. Also the overall dependence of the conductivity of the three-dimensional continuum percolation model on the concentration of the insulating phase, enhanced interface conductance and particle size, was in good qualitative agreement with experiments. Recently, the same model has been used [10] for studying the ionic transport in nano- and microcrystalline composites with thermally activated component conductivities. It was found that the ionic conductivity depends drastically on explicitly taking into account the two different grain sizes of both components. The above mentioned examples show the real importance of incorporation of the changes in the size of the dispersed particles for effective media theories, which describe at least qualitatively the conduction properties.

---

[*] e-mail: piaser@uni.opole.pl

A novel, so called local linearization method of determination of the effective nonlinear conductivity has been given recently [11]. Also the network extension [12] of EMA embodies essential physics of macroscopic properties of random heterogeneous materials. Within such an approach, the macroscopic nonlinear response characteristics of metal-insulator mixtures was obtained for a minimal model bond percolative network [13] with three types of bonds: ohmic, tunnelling and purely insulating ones. In Ref.[2], for certain GSDs the non-monotonous critical behaviour of conductor-insulator and conductor-superconductor systems was demonstrated. Within this approach every grain for H-(L-)phase of high $\sigma^H$ (low $\sigma^L$) conductivity is composed of grains of type '1' $\equiv s_1$, which can 'occupy' alone the centre of a unit bond on a reference square lattice marking a pure unit H-(L-)bond. The effective dc conductivity $\sigma^*(p)$, where $p$ is the fraction of unit H-bonds, was considered on a coarsened lattice for systems with grains simple in form only. Here we study a modified two-dimensional model that allows for differently shaped grains and complex GSDs. A main goal is to find how both GSD(H) and GSD(L) and their reverse combination affect $\sigma^*(p)$. This provides insight into how empirical size histogram information can be used to predict effective properties, e.g. for random granular films and their dual counterparts.

**2. Model and results**

Consider now the coarsened square lattice with model bonds each of length $l > 1$. To each of the bonds a square model cell is assigned consisting of $l^2$ 'positions', that is, the centres of neighbouring unit bonds. Now, besides square shape, e.g., for '4' $\equiv s_4$ grains, one allows also rod-like (if $l \geq 4$) or T-like grains, letter L or 'zigzag' shaped grains and their mirror reflections. In Fig. 1, exemplary configurations at length scale $l = 3$ are presented for certain GSDs. Given that the area of an $s_n$ grain is 'measured' by the number $n$ of the occupied positions, a simple rule (analogous to the 'shaking' procedure) used in numerical simulations reads: the sum of all H- and L-grain areas inside of every model cell must be equal to $l^2$. Let $N$ and $N^K(p)$ for a given $p$ be, respectively, the number of all model bonds and bonds of type $K = H, L,$ and $M$. We comment further on mixed case below. The bond fractions defined as $F_K(p) \equiv N^K(p)/N$ fulfil the obvious condition

$$F_H(p) + F_M(p) + F_L(p) = 1. \tag{1}$$

Similarly to Ref.[2], $\gamma_i \equiv n_i/l^2$ and $1 - \gamma_i$ is the local fraction of $n_i$ and $l^2 - n_i$ of unit H- and L-bonds in $i$-th model cell. Thus, for a pure H-(L-)bond we have $\gamma_i = 1$ (= 0), while for $i$-th M-bond $0 < \gamma_i < 1$. Now, for the fraction of unit H-bonds the elementary relation holds

$$p = \frac{1}{N}[N^H(p) + \sum_{i\in\{M\}}\gamma_i] \equiv F_H(p) + \gamma(p)F_M(p), \qquad (2)$$

where $\gamma(p)$ is an average fraction of unit H-bonds per mixed cell.

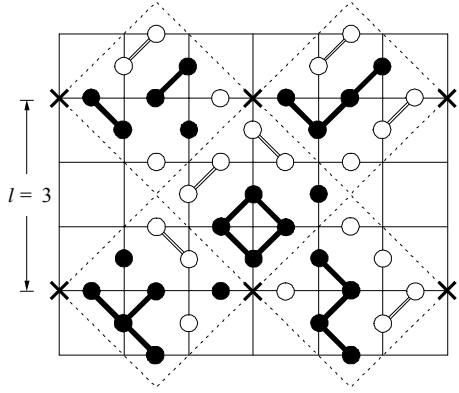

Fig. 1. Examples of the model bonds (×——×) each of length $l=3$ and cells (marked by dashed lines), each consisting of $l^2=9$ positions for chosen distributions $\{s_1, s_2, s_4\}:\{s_1, s_2\}$ of H-phase (filled circles) and L-phase (open circles), respectively. In contrast to the previously used model [2], differently shaped grains (here of $s_4$ area) are also allowed. Note that the presented model bonds (and cells) belong to the mixed type, $M$.

In order to simplify the problem the following approximation is made: each mixed-cells with a certain $\sigma_i^M$ conductivity can be replaced by a representative mixed cell with $\gamma(p)$ and $1-\gamma(p)$ fractions of H- and L-bonds, and the mixed $\sigma^M(p)$ conductivity be calculated within the EMA from a corresponding quadratic expression, $(\sigma^M)^2 + \sigma^M(\sigma^H - \sigma^L)(1-2\gamma(p)) - \sigma^H\sigma^L = 0$. Thus,

$$\sigma^M(p) = \left[-(\sigma^H - \sigma^L)(1-2\gamma(p)) + \sqrt{(\sigma^H - \sigma^L)^2(1-2\gamma(p))^2 + 4\sigma^H\sigma^L}\right]/2. \qquad (3)$$

The general distribution of the corresponding bond conductivities $\sigma^L < \sigma^M(p) < \sigma^H$ is now of a simpler form compared to Eq.(1) in Ref.[2] (on page 405, it should read without a misprinted '$= 0$' on r.h.s.),

$$P(\sigma) = F_H(p)\,\delta(\sigma - \sigma^H) + F_M(p)\,\delta(\sigma - \sigma^M(p)) + F_L(p)\,\delta(\sigma - \sigma^L). \qquad (4)$$

Following [12] the equation for the overall effective conductivity $\sigma^*(p)$ can now be written as

$$F_H(p)\frac{\sigma^H - \sigma^*(p)}{\sigma^H + \sigma^*(p)} + F_M(p)\frac{\sigma^M(p) - \sigma^*(p)}{\sigma^M(p) + \sigma^*(p)} + F_L(p)\frac{\sigma^L - \sigma^*(p)}{\sigma^L + \sigma^*(p)} = 0. \qquad (5)$$

The fractions $F_H(p)$, $F_M(p)$, and $F_L(p)$ are difficult to calculate analytically due to the large number of possible local configurations in a model cell. The same refers to the $\gamma(p)$ function. However, using Eqs.(1) and (2) at least two of the four quantities can be treated as independent ones, for instance, $F_M(p)$ and $\gamma(p)$. Thus, from Eqs.(1) and (2) we have

$$F_H(p) = p - \gamma(p)F_M(p),$$
$$F_L(p) = (1-p) - (1-\gamma(p))F_M(p). \qquad (6)$$

As observed in numerical simulations for various GSDs, the fractions $F_H(p)$ and $F_L(p)$ are concave functions while the mixed fraction $F_M(p)$ is a convex one reaching a maximum, say $\beta$, at certain $p=\alpha$. The topological equivalence of equally sized H- and L-grains implies $\alpha=0.5$ while for complex GSDs we expect $\alpha\neq 0.5$. Supported by preliminary computer simulations, a trial function approximating the behaviour of $F_M(p)$ is proposed as

$$F_M(p) \approx \begin{cases} a_0 + a_1 p + a_2 p^2 & \text{for } 0 < p < \alpha, \\ \beta & \text{for } p = \alpha, \\ b_0 + b_1 p + b_2 p^2 & \text{for } \alpha < p < 1, \end{cases} \qquad (7)$$

where constant coefficients $a_k$ and $b_k$ can be determined from the conditions

$$\begin{aligned} &(i) \quad \lim_{p \to 0^+} F_M(p) = 0 \quad \text{and} \quad \lim_{p \to 1^-} F_M(p) = 0, \\ &(ii) \quad \lim_{p \to \alpha^-} F_M(p) = \lim_{p \to \alpha^+} F_M(p) = \beta, \\ &(iii) \quad \lim_{p \to \alpha^-} \frac{dF_M(p)}{dp} = \lim_{p \to \alpha^+} \frac{dF_M(p)}{dp} = 0. \end{aligned} \qquad (8)$$

The final form is

$$F_M(p) = \begin{cases} \left[1 - \left(\frac{\alpha - p}{\alpha}\right)^2\right]\beta \equiv [2\alpha p - p^2]\beta/\alpha^2 & \text{for } 0 < p \le \alpha, \\ \left[1 - \left(\frac{p - \alpha}{1 - \alpha}\right)^2\right]\beta \equiv [1 - 2\alpha + 2\alpha p - p^2]\beta/(1-\alpha)^2 & \text{for } \alpha \le p < 1. \end{cases} \qquad (9)$$

From Eqs. (6) and (9) the remaining $F_H(p)$ and $F_L(p)$ bond fractions can be also parameterized by $\alpha$, $\beta$, and $\gamma(p)$. Note that the final form of the model $F_M(p)$ function possesses the desirable symmetry property (i) $F_M(p, \alpha, \beta) = F_M(1-p, 1-\alpha, \beta)$. As computer simulations revealed, such a symmetry is connected with the phase interchanging. So, the more detailed form is $F_M(\boldsymbol{p}, \sigma^H; 1-p, \sigma^L; \alpha, \beta) = F_M(p, \sigma^L; \boldsymbol{1-p}, \sigma^H; 1-\alpha, \beta)$, where the bold symbols clearly show that in the simplified notation both the concentrations $p$ and $1-p$ refer to the H-phase. Remembering this, in the following the simpler notation will be used. Further symmetry relations are: (ii) $\gamma(p) + \gamma(1-p) = 1$, (iii) $F_H(p, \alpha, \beta, \gamma(p)) = F_L(1-p, 1-\alpha, \beta, 1-\gamma(p))$ and vice versa, see Eqs. (6). Further, Eq. (3) for the mixed conductivity $\sigma^M(p)$ as well as Eq. (5) for the effective conductivity $\sigma^*(p)$ is invariant under the simultaneous interchange $\gamma(p) \leftrightarrow 1 - \gamma(p)$, $\sigma^H \leftrightarrow \sigma^L$. A general identity linking the two-dimensional effective conductivity of a two-phase macroscopically isotropic system to the effective conductivity of the same microstructure but with the phases interchanged, called the phase-interchange relation, has been obtained by Straley [14]. The corresponding expression is

$$\sigma^*(p, \sigma^H; 1-p, \sigma^L; \alpha, \beta, \gamma(p))\, \sigma^*(p, \sigma^L; 1-p, \sigma^H; 1-\alpha, \beta, 1-\gamma(p)) = \sigma^H \sigma^L. \qquad (10)$$

This relation was checked making use of computer simulations. For example, see Fig. 2a, where the exemplary pair of $p = 0.2$ and $1-p = 0.8$ ($p = 0.6$ and $1-p = 0.4$) shows that the sum of logarithms of the corresponding $\sigma^*$ values indicated by black (white) arrows always equals $-3$, as expected for the given phase conductivities, $\sigma^H = 1$ and $\sigma^L = 10^{-3}$ in a.u.

Let us check the limiting behaviour of our model. Its examination shows that for $l = 1$ there are no mixed model cells, $\gamma(p) = 0$ or $1$, $F_M(p) = 0$, $F_H(p) = p$, $F_L(p) = 1-p$, and the Bruggeman model with a standard binary distribution of pure bonds is recovered. For $l \gg 1$ and medium $p$ there is a negligible probability of appearance of pure model cells, so $\gamma(p) \approx p$, $F_M(p) \approx 1$, $F_H(p) \approx 0$, $F_L(p) \approx 0$, and from Eq. (5) results that $\sigma^M(p) \approx \sigma^*(p)$, i.e., the mixed conductivity becomes the global one, as expected. If both phases consist of equally mono-sized grains and the system is considered at larger scales, then using an auxiliary parameter, e.g. $\delta \equiv F'_H|_{p=\alpha}$, where the derivative is taken over $p$, a linear function for an approximated description of $\chi(p)$ can be deduced, $\chi(p) \approx (\beta + \delta - 1)/2\beta + p(1 - \delta)/\beta$. In general, however, the $\chi(p)$ values for a given concentration are evaluated by performing computer simulations.

At this stage, we would like to concentrate on the numerical evaluation of model functions $\chi(p)$ and $F_M(p)$ to calculate $\sigma^M(p)$. Also $F_H(p)$ and $F_L(p)$ values are evaluated and then Eq. (5) is solved for the most interesting range of length scales, i.e., when $l$ is comparable with the largest grain size of the given GSD(H) and GSD(L). Computer simulations on a square lattice with 4900 bonds each of length $l=6$ are performed for $p=0.1, 0.2, ..., 0.9$. The probability of selecting a H-(L-grain) is proportional to the initial area fraction of grains of a given type. For the considered concentrations this rule is good enough to overpass a blockage problem and it is not so restrictive for randomness as, e.g., the sequential drawing. The effective conductivity is averaged over 1000 run trials. With a high accuracy the same results are obtained by first averaging the bond fractions and then solving Eq. (5). To illustrate how the effective conductivity depends on the respective grain sizes we consider first H-phase log-normal distributed among the mono-sized grains of the L-phase for two different distributions. Then, a slightly perturbed log-normal GSD(H) and mono-sized L-grains as well as the arbitrarily chosen GSD(L) are considered. Simultaneously, the corresponding phase-interchanged situations are also illustrated in Figs. 2 and 3.

For commercially available metal powders used in metal oxide/metal composites, see for instance Ref. [15], the particle diameters $d$ can be characterized by a log-normal distribution

$$f(d) = \frac{1}{d\sigma\sqrt{2\pi}} \exp\left[-(\ln d - \mu)^2/2\sigma^2\right], \tag{11}$$

where $\mu$ is the average of the logarithms of the diameters and $\sigma$ is the standard deviation of the distribution. Assuming that the side lengths $d$ of 'square' grains satisfy Eq. (11) and using the equality $d^2 = s_n$, this formula can be rewritten to account for the grain area distribution $dN_s = f(s)ds$,

$$f(s) = \frac{A}{s\sigma_s\sqrt{2\pi}} \exp\left[-(\ln\sqrt{s} - \mu_s)^2/2\sigma_s^2\right], \tag{12}$$

where $\mu_s$ is the average of the logarithms of the grain areas and $\sigma_s$ is the standard deviation of the distribution. Because we consider discrete values of the areas of grains $s_n$, a factor $A$ allows a set of possible values of corresponding relative area frequencies to be constrained by $\Sigma f_i = 1$.

Let us consider two $\{s_4, s_9, s_{16}, s_{25}, s_{36}\}$ distributions for H-grains with the relative area frequencies $f_i(1)$ ($f_i(2)$) $\approx$ 0.007 (0.138), 0.204 (0.551), 0.417 (0.250), 0.272 (0.051), 0.100 (0.010), depicted in the inset in Fig. 2a, filled (open) diamonds, and common $\{s_1\}$ uniform distribution for L-grains. The log-normal distribution curves have the same $\sigma_s = 0.24$, but different $\mu_s(1) = 1.5$ and $\mu_s(2) = 1.2$. The average $\sigma^*(p)$ values are shown in Fig. 2a, (filled and open) diamonds. The dashed (solid) lines 1 (2) are guides to the eye only.

For both log-normal distributions considered the figure shows, that H-grains polydispersed in size when embedded into a 'matrix' of mono-sized L-grains, favour a higher effective conductivity in comparison to the reverse case. Note that for these distributions with a fixed $\sigma_s$ the higher increase in $\sigma^*(p)$, referring to the 'transition' (1-r) $\rightarrow$ (1), is connected with the higher average of the logarithms of the grain areas. So, the frequency $f_i(1)$ with a relatively higher number of larger H-grains is more effective in this context than $f_i(2)$. A hysteresis-loop like behaviour of the effective conductivity, see Figs. 2a and 3a, strikingly remains the behaviour of conductivity obtained for the two-dimensional cooper-graphite dual media, see Figs. 1 and 2 in Ref. [16]. There, the former figure shows examples of so called NEG (negative board) and POS (positive board) patterns. In the latter figure the experimental points of conductivity versus graphite area fraction are fitted by a general effective medium (GEM) equation, which combines most aspects of percolation and effective medium theories [17].

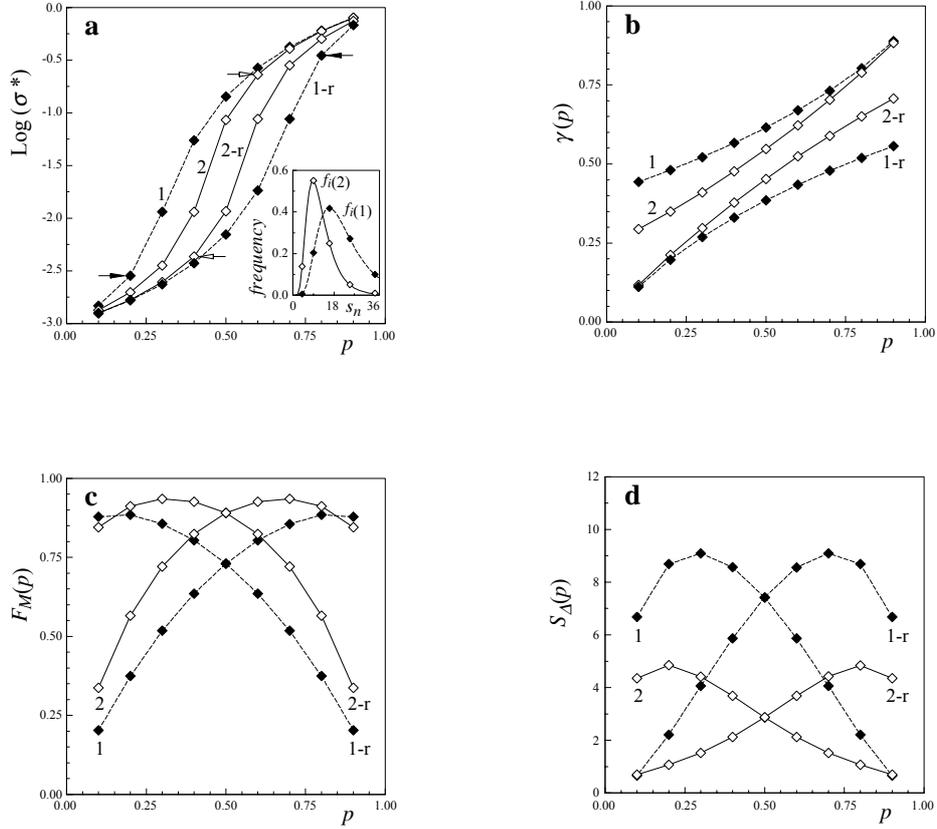

Fig. 2. Examples of GSD dependence of the model related quantities for $p$ = 0.1, 0.2, ..., 0.9 and for $\sigma^H$ = 1 and $\sigma^L = 10^{-3}$ in a.u. at length scale $l$ = 6. a) The effective conductivity $\log(\sigma^*(p))$ for case (1) with $\{f_i(1;H)\}:\{s_1(L)\}$ (filled diamonds and top dashed line), for case (1-r) with $\{s_1(H)\}:\{f_i(1;L)\}$ (filled diamonds and bottom dashed line), for case (2) with $\{f_i(2;H)\}:\{s_1(L)\}$ and (2-r) with $\{s_1(H)\}:\{f_i(2;L)\}$ (open diamonds and middle solid lines). Black and white arrows illustrate the phase-interchange relation (10). Area frequencies $f_i(1)$ (filled diamonds) and $f_i(2)$ (open diamonds) are shown in the inset. b) The corresponding $\gamma(p)$ values. c) The corresponding fractions of mixed bonds $F_M(p)$. d) The spatial inhomogeneity of coarsened lattice quantified by entropic measure $S_\Delta(p)$ in the thermodynamic limit [18, 20].

In Fig. 2(b) the corresponding $\gamma(p)$ functions are ordered from the top to the bottom, in the same sequence as the conductivity curves. The inversion symmetry of the appropriate pairs of $\gamma(p)$ functions with respect to the point $p$ = 0.5, $\gamma$ = 0.5 appear in agreement with the already mentioned symmetry relation (ii). The corresponding mixed-bond fractions with the reflection symmetry under the replacement $p \leftrightarrow 1-p$ described in relation (i) are presented in Fig. 2c. Now, for most $p$ values the following tendency appears for the two log-normal distributions: the lower mixed bonds fraction $F_M(1) < F_M(2)$ favours the higher effective conductivity $\sigma^*(1) > \sigma^*(2)$, while for the reverse cases the opposite connection is observed, namely $F_M$(1-r) < $F_M$(2-r) and $\sigma^*$(1-r) < $\sigma^*$(2-r). Consider now the quantitative evaluation of microstructure attributes for searching their possible connections with the macroscopic properties of a disordered system. The quantity of interest is the entropic measure of the length scale dependent spatial inhomogeneity for systems of finite-sized objects, which was fully described and applied in recent papers [18−20]. Using the notation appropriate for the present work, the measure can be written (at fixed length scale $l$) in the thermodynamic limit as a function of concentration,

$$S_\Delta(p) = -l^2 \{ p \ln p + (1-p) \ln(1-p) \\ - F_M(p)[\gamma(p) \ln \gamma(p) + (1-\gamma(p)) \ln(1-\gamma(p))] \}. \tag{13}$$

For a given $p$ this measure quantifies the average (per model cell) deviation of a general configurational macrostate (related to the actual system's configuration) from the reference macrostate (related to the most spatially uniform distribution). Formula (13) includes combinations of both $\gamma(p)$ and $F_M(p)$ functions. However, taking into account relations (i) and (ii), the measure shows a symmetry of the latter one, $S_\Delta(p, \alpha, \beta) = S_\Delta(1-p, 1-\alpha, \beta)$. Now, the higher spatial inhomogeneity $S_\Delta(1) > S_\Delta(2)$, see Fig. 2d, favours the higher effective conductivity $\sigma^*(1) > \sigma^*(2)$, while for the reverse cases we have the opposite relation, namely $S_\Delta$(1-r) > $S_\Delta$(2-r) and $\sigma^*$(1-r) < $\sigma^*$(2-r). Note the asymmetry of the $S_\Delta(p)$ curves regarding $p = 0.5$, which confirms the topological non-equivalence of H- and L-phases [20].

Let us now consider a perturbed log-normal $\{s_4, s_9, s_{16}, s_{25}, s_{36}\}$ distribution denoted as (I), with the slightly deviated (in comparison to the previous case (1)) frequencies $f_i'(1;H) \approx$ 0.0463, 0.1531, 0.4535, 0.2268, 0.1204, depicted in the inset in Fig. 3a (filled triangles), and $\{s_1\}$ uniform distribution of L-grains. Next, we replace $\{s_1(L)\}$ by $\{s_1, s_2, s_4, s_{12}, s_{36}\}$ distribution with arbitrarily selected relative area frequencies $g_i(L) \approx$ 0.2653, 0.1361, 0.0680, 0.0204, 0.5102 depicted in the inset in Fig. 3a (open triangles). This way we obtain a modification of case (I) denoted here as (mI). The two reverse cases with the phases mutually interchanged are marked as (I-r) and (mI-r). Now, the clear increase of $\sigma^*(p)$ with regard to case (I) appears in Fig. 3a (filled circles). Quite surprisingly, the presence of the largest $s_{36}$ L-grains seem to be necessary for that. For the reverse case (mI-r) the largest reduction of $\sigma^*(p)$ can be also seen in Fig. 3a (open circles).

The dependence of $\sigma^*(p)$ on GSDs can be roughly understood on the basis of Figs. 3b and c. Let the arrows ↑ and ↓ symbolise increasing and decreasing values of the quantities followed by them and, for a given $p$, consider the exemplary 'transition': (I) → (mI). This implies $\gamma$↑ (see Fig. 3b), $F_H$↑ (not shown here), $F_M$↓ (see Fig. 3c), $F_L$↑ (not shown here), $\sigma^M$↑ (not shown here), and finally $\sigma^*$↑ (see Fig. 3a). The decrease of $F_M(p)$ induces the favourable higher $\sigma^*(p)$ increase of $F_H(p)$ that is more significant than the unfavourable increase of $F_L(p)$. On the other hand, the increase of the $\gamma(p)$ function is also essential for the increase of $\sigma^M(p)$, particularly for the middle concentration values. All these effects lead finally to a higher $\sigma^*(p)$. This kind of sensitivity to various GSDs is a peculiar feature of our model in which spatially distributed mixed cells with various conductivities $\sigma_i^M$ are replaced by the identically distributed 'average' mixed cells, each of conductivity $\sigma^M(p)$. It is worth to mention that a quite intriguing similarity appears between the $\gamma(p)$ behaviour shown in Fig. 3b, e.g. for the mono-sized cases with $\{s_9(H)\}:\{s_9(L)\}$, symbols '+', and with $\{s_6(H)\}:\{s_6(L)\}$, open squares, and the so called microstructure parameter $\zeta_1$ (see Table 1 and Fig. 1 in Ref. [21]). A similar observation refers also to Fig. 2b.

Continuing the comparison of cases (I) with (mI) and (I-r) with (mI-r) we must remember that both modifications were done by replacing $\{s_1\}$ with $\{s_1, s_2, s_4, s_{12}, s_{36}\}$ distribution. Due to the reduced $F_M(p)$ for both cases and the proper behaviour of the $\gamma(p)$ function we observe for the two modifications a higher spatial inhomogeneity than for the initial situations, see Fig. 3d.

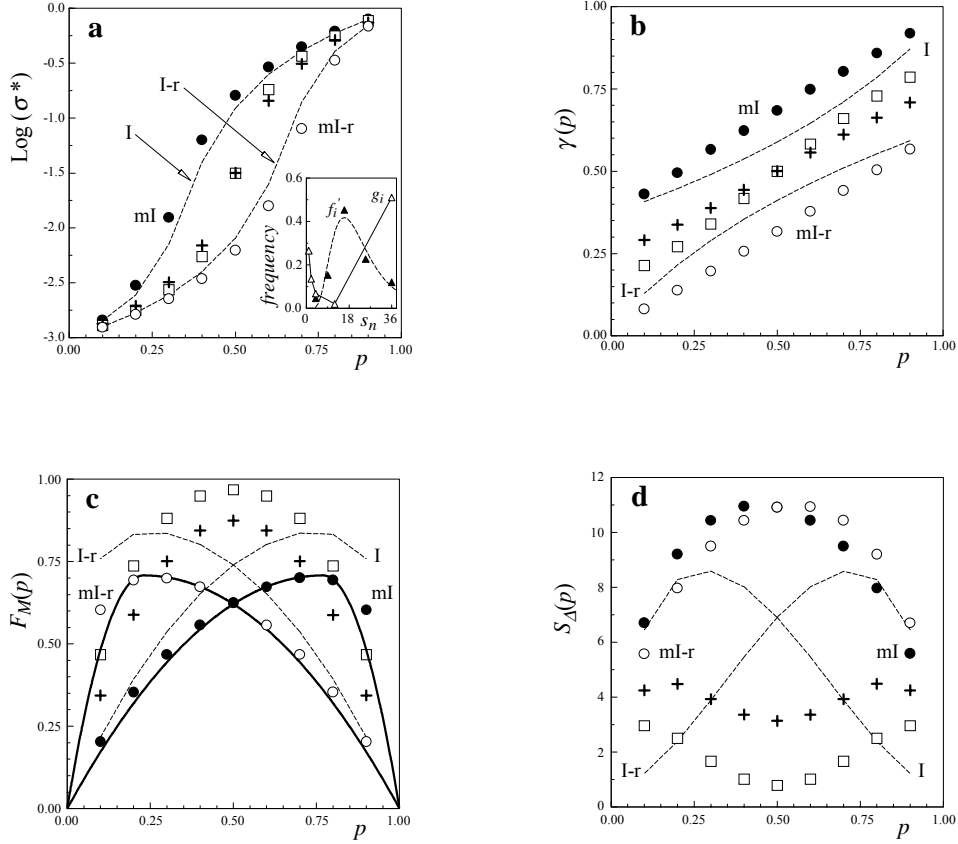

Fig. 3. Further examples of GSD dependence for the model related quantities with the same concentrations as in Fig. 2, pure phase conductivities and length scale. a) The effective conductivity $\log(\sigma^*(p))$ for case (I) with $\{f_i{'}(H)\}:\{s_1(L)\}$ (top dashed line), for case (I-r) with $\{s_1(H)\}:\{f_i{'}(L)\}$ (bottom dashed line), for the modified case (mI) with $\{f_i{'}(H)\}:\{g_i(L)\}$ (filled circles), for the reverse modified case (mI-r) with $\{g_i(H)\}:\{f_i{'}(L)\}$ (open circles), for mono-sized cases with $\{s_9(H)\}:\{s_9(L)\}$ (symbols '+') and with $\{s_6(H)\}:\{s_6(L)\}$ (open squares). In the inset the area frequencies are shown, $f_i(1)$ (dashed line), its perturbed version $f_i{'}$ (black triangles), and specifically chosen $g_i$ (open triangles). b) The corresponding $\gamma(p)$ values. c) The corresponding fractions of mixed bonds $F_M(p)$ with an analytical result Eq.(9) for both modified cases with estimated $\alpha \approx 0.77$ for (mI) and $\alpha \approx 0.23$ for (mI-r) and common $\beta \approx 0.71$, thick solid lines. d) The corresponding spatial inhomogeneity of coarsened lattice.

The figure also shows the asymmetry of the $S_\Delta(p)$ curves regarding $p = 0.5$, which again confirms the topological non-equivalence of polydispersed H- and L-phases in contrast to the two symmetrical mono-sized cases. Topologically non-equivalent microstructure of a system means that the two phases are not ideally randomly dispersed and deviations from randomness exist because of local ordering correlations. We can see again that the higher spatial inhomogeneity $S_\Delta(mI) > S_\Delta(I)$ is connected with the higher effective conductivity $\sigma^*(mI) > \sigma^*(I)$, and for the reverse cases the opposite relation $S_\Delta(mI\text{-}r) > S_\Delta(I\text{-}r)$ and $\sigma^*(mI\text{-}r) < \sigma^*(I\text{-}r)$ appears. Such a correspondence is found in both investigated cases.

In general however, the relations between $S_\Delta(p)$ and $\sigma^*(p)$ are more complicated. Qualitatively different behaviour can be seen when the two different mono-sized cases '+' and '□' are compared. Now, $\chi(p; +) = \chi(p; □) = 0.5$ for $p = \alpha = 0.5$ and the relation between spatial inhomogeneity and effective conductivity changes for the reverse one when the concentration passes this $\alpha$ value, although the spatial inhomogeneity $S_\Delta(+) > S_\Delta(□)$ for every $p$. The $\chi(p)$ function provides the simplest way to predict what the sequence of $\sigma^*(p)$ curves is. For a given $p$ and H-phase log-normal distributed among mono-sized or polydispersed grains of L-phase, simulation results show that the larger $\chi(p)$ values correlate with higher effective conductivity, compare Figs. 2a and b and Figs. 3a and b.

Another common consequence of polydispersity in grain sizes is a blockage problem for low and high concentrations, usually for $p < 0.1$ and $p > 0.9$, linked also with a type of GSD. However, the model can be readily extended to one containing pores. Then, a conductance equal to zero is attributed to each unoccupied unit bond. Such purely insulating unit bonds can be present only in mixed cells of our model. Denoting by $\chi_0(p)$ its average fraction per mixed cell, the corresponding quadratic expression for the mixed $\sigma^M(p)$ conductivity is now $(\sigma^M)^2 + \sigma^M[(\sigma^H - \sigma^L)(1 - 2\chi(p)) + 2\sigma^L\chi_0(p)] - \sigma^H\sigma^L(1 - 2\chi_0(p)) = 0$. Some symmetry properties of the model functions are lost but the blocking is practically eliminated.

One more remark is in order. Concerning the question of extending the model ability to three dimensions, the answer is positive for all odd lengths of model bonds, $l = 2k + 1$, and for those even lengths that are of $l = 4k$ form, where $k$ is a natural number. For example, for $l = 3$ the model cell of the coarsened regular lattice consists of 19 unit bonds. All we need is replacing the basic conductivity formula within EMA by its well-known three-dimensional counterpart.

### 3. Conclusions

In this paper, the two-dimensional lattice model describing, within EMA, the concentration dependence of the effective conductivity for binary disordered systems has been examined. Apart from the simplicity of the model, a wide variety of grain areas and shapes is included. Such a model can be also applied to artificially synthesised composite systems with different shapes of grains of both phases. Its analytical form allows for consideration of the whole three-parameter space, $0 < \alpha, \beta, \gamma < 1$. The model is highly sensitive to a type of GSDs as well as their mutual interchanging. In such cases it behaves in a way typical for dual media. We show that there exists a simple correlation between the model cell average concentration of the H-component and the effective conductivity. The topological non-equivalence of the system's phases can be easily detected by the entropic measure of spatial inhomogeneity of the model cells. Incorporation of the porosity of disordered media is straightforward in this model. Finally, it should be stressed that the model possesses a characteristic feature. Namely, it allows for a description of the effective conductivity from different size scales point of view. Thus, it may exhibit information not necessarily revealed by other models of disordered media.

**References**

[1] S. TORQUATO, Random Heterogeneous Materials, Springer-Verlag New York 2002.
[2] R. PIASECKI, phys. stat. sol. (b) **209**, 403 (1998).
[3] S. TOYOFUKU, T. ODAGAKI, Mat. Sci. Eng. A **217/218**, 381 (1996).
[4] S. TOYOFUKU, T. ODAGAKI, J. Phys. Soc. Jap. **66**, 3512 (1997).
[5] R.E. AMRITKAR, MANOJIT ROY, Phys. Rev. E **57**, 1269 (1998).
[6] DAE GON HAN, GYEONG MAN CHOI, Solid State Ionics **106**, 71 (1998).
[7] I.C. RAKENBURG, R.J. ZIEVE, Phys. Rev. E **63**, 061303 (2001).

[8] H.E. S. ROMAN, M. YUSSOUFF, Phys. Rev. B **36**, 7285 (1987).
[9] H.E. S. ROMAN, J. Phys.: Condens. Matter **2**, 3909 (1990).
[10] S. INDRIS, P. HEITJANS, H.E. ROMAN, A. BUNDE, Phys. Rev. Lett. **84**, 2889 (2000).
[11] A.A. SNARSKII, M. ZHENIROVSKIY, Physica B **322**, 84 (2002).
[12] S. KIRKPATRICK, Rev. Mod. Phys. **45**, 574 (1973).
[13] ABHIJIT KAR GUPTA, ASOK K. SEN, Phys. Rev. B **57**, 3375 (1998).
[14] J.P. STRALEY, Phys. Rev. B **15**, 5733 (1977).
[15] J. KIEFER, J.B. WAGNER, J. Elecktrochem. Soc. **135**, 198 (1988).
[16] D.S. McLACHLAN, J. Phys. C: Solid State Phys. **21**, 1521 (1988).
[17] D.S. McLACHLAN, J. Phys. C: Solid State Phys. **20**, 865 (1987).
[18] R. PIASECKI, Physica A **277**, 157 (2000).
[19] R. PIASECKI, Surf. Sci. **454–456**, 1058 (2000).
[20] R. PIASECKI, A. CZAIŃSKI, phys. stat. sol. (b) **217**, R10 (2000).
[21] A.P. ROBERTS, M. TEUBNER, Phys. Rev. E **51**, 4141 (1995).